\documentstyle[12pt,epsfig,axodraw]{article}
\def\unity{{\hbox{1\kern-.8mm l}}}
\def\dfrac#1#2{{\displaystyle\frac{#1}{#2}}}
\def\be{\begin{equation}}
\def\ee{\end{equation}}
\def\bc{\begin{center}}
\def\ec{\end{center}}
\def\bea{\begin{eqnarray}}
\def\eea{\end{eqnarray}}

\def\cl{{\cal L}}

\def\co{{\cal O}}
\def\gol{\tilde{G}}

\def\ga{\gamma}
\def\la{\lambda}
\def\lab{\ov\lambda}
\def\gf{G_{\small F}} 
\def\fb{\ov{f}}
\def\fc{f^c}
\def\fcb{\ov\fc}
\def\ft{\tilde{f}}
\def\fct{\tilde{\fc}}
\def\fl{\tilde{f}_L}
\def\fr{\tilde{f}_R}
\def\lt{\tilde{\ell}}

\def\mt{\tilde{m}}
\def\mmt{{\bf\mt}}
\def\mll{\tilde{m}^2_{LL}}
\def\mrr{\tilde{m}^2_{RR}}
\def\mlr{\tilde{m}^2_{LR}}
\def\mrl{\tilde{m}^2_{RL}}
\def\th{\theta}

\def\pz{\tilde{G}}
\def\pzb{{\ov{\tilde{G}}}}
\def\dmu{\partial^\mu}
\def\dmd{\partial_\mu}

\def\dnd{\partial_\nu}

\def\lsq{{\Lambda^2}}
\def\ov{\overline}
\def\smd{\sigma_{\mu}}
\def\smu{\sigma^{\mu}}
\def\smdb{{\bar\sigma}_{\mu}}
\def\smub{{\bar\sigma}^{\mu}}

\def\snu{\sigma^{\nu}}

\def\snub{{\bar\sigma}^{\nu}}
\def\smn{\sigma^{\mu\nu}}
\def\smnb{{\bar\sigma}^{\mu\nu}}
\def\simlt{\stackrel{<}{{}_\sim}}

\def\to{\rightarrow}
\begin{document}
\begin{titlepage}
\vspace*{-1cm}
\phantom{hep-th/0006036} 
\hfill{DFPD-00/TH/28}
\vskip 2.0cm
\begin{center}
{\Large\bf Flavour non-conservation in goldstino interactions} 
\end{center}
\vskip 1.5  cm
\begin{center}
{ {\large 
Andrea Brignole\footnote{E-mail address: 
brignole@padova.infn.it}
and 
Anna Rossi\footnote{E-mail address: 
arossi@padova.infn.it}}
\\
\vskip .5cm
{\it Istituto Nazionale di Fisica Nucleare, Sezione di Padova, 
\\
Dipartimento di Fisica `G.~Galilei', Universit\`a di Padova, 
\\
Via Marzolo n.8, I-35131 Padua, Italy}}
\end{center}
\vskip 3.0cm
\begin{abstract}
\noindent
We point out that the interactions of goldstinos with 
matter supermultiplets are a potential source of 
flavour violation, if fermion and sfermion mass matrices 
are not aligned and supersymmetry is spontaneously 
broken at a low scale. 
We study the impact of those couplings on low-energy 
processes such as $\mu \to e \ga$, $\mu \to eee$,
$K \to \mu^+ \mu^-$, $K$-$\ov{K}$ transitions
and analogous ones.
Moreover, we address the issue of flavour violation
in low-energy processes involving  two goldstinos 
and two matter fermions, generalizing earlier 
results obtained in the flavour-conserving case.  
\end{abstract}
\end{titlepage}
\setcounter{footnote}{0}
\vskip2truecm


\vspace{1 cm}

\noindent
{\large \bf 1. Introduction}
\vspace{0.3 cm}

Contributions to flavour changing neutral currents (FCNC) 
are adequately suppressed in the standard model (SM) \cite{fcsm}.
Supersymmetric extensions of the SM generate additional
contributions to FCNC, even in the case of minimal field 
content and conserved R-parity (MSSM) \cite{fcmssm}. Such effects
are due to both charged and neutral couplings.
In particular, the interactions of matter multiplets with 
neutral gauginos, which have the form $g \ft^* f \lambda$,
are a potential source of flavour violation that has 
no counterpart in the SM.
Indeed, if fermion and sfermion mass matrices are not
diagonal in the same superfield basis, those couplings
induce flavour changing effects, which have been 
extensively studied (see e.g. \cite{gab,ggms} and references 
therein). On the other hand, if fermion and sfermion mass
matrices are misaligned, FCNC receive additional contributions 
from another class of neutral couplings: the interactions of 
matter multiplets with goldstinos. This observation is the 
starting point of the investigation we would like to 
present in this paper.

We recall that the supersymmetry breaking masses of the MSSM 
are expected to originate from the spontaneous breaking of 
supersymmetry in some underlying theory. 
This phenomenon entails the existence of a goldstone fermion, 
the goldstino, which couples to matter and gauge supermultiplets
in a characteristic way. For instance, the interaction 
of a goldstino $\gol$ with a fermion-sfermion pair has the form 
$(\Delta m^2 /F) \,  \tilde{f}^* f \gol$, where $\sqrt{F}$ is 
the supersymmetry breaking scale and $\Delta m^2$ denotes the 
mass splitting in the sector under consideration \cite{fayet1}. 
The form of the interaction resembles that of a neutral gaugino,
although the strength is $\Delta m^2/F$ instead of $g$. 
If the mass splitting has electroweak size whilst $\sqrt{F}$ is 
much larger, the goldstino is essentially decoupled.
Here we are interested in the opposite scenario, in which
$\sqrt{F}$ is not far from the Fermi scale $1/\sqrt{\gf}$. 
In this case the interactions involving goldstinos are no longer 
negligible and can have observable effects\footnote{In the 
context under consideration, the 
words goldstino and light gravitino can be interchanged, thanks 
to the equivalence theorem \cite{fayet1,cddfg}. We recall that 
the gravitino becomes massive
by absorbing the goldstino. Its mass is related to the supersymmetry 
breaking and Planck scales as $m_{3/2}\! =\! F/(\sqrt{3}M_P)$. For 
$F \simeq \co(G_F^{-1})$, $m_{3/2} \simeq \co(10^{-5})\, {\rm eV}$.}. 
How to obtain low values 
of $\sqrt{F}$ in concrete models is an open issue, which we will not 
discuss.
We only recall that such a possibility is not ruled out by present 
experiments, as shown for instance by recent studies on goldstino 
pair-production at $e^+ e^-$ and hadron colliders \cite{coll}. 
Therefore we believe that it is worth exploring this scenario 
also in connection to flavour changing
phenomena, taking into account the possible flavour structure 
of sfermion mass matrices (that is, of $\Delta m^2$).
This is the purpose of the present paper. 
Notice that we do not try to `solve' the supersymmetric
flavour problem, in contrast to more fundamental approaches which
directly address the origin of flavour and/or supersymmetry
breaking, or at least the mediation of that breaking. For instance,
in models in which such mediation is due to ordinary gauge interactions 
(for a review, see e.g. \cite{gr}) $\sqrt{F}$ is relatively low, but
still two or three orders of magnitude larger than $1/\sqrt{G_F}$, so 
the goldstino is very weakly coupled to matter; furthermore, the 
mediation mechanism generates flavour-blind sfermion masses, so flavour 
violations through gaugino couplings are automatically suppressed
in those models. 
Our approach here is somewhat orthogonal, i.e. phenomenological
rather than model-based: we would like to 
study the flavour dependence of goldstino-matter interactions 
and the related implications for flavour changing processes
in a general way, by treating $\sqrt{F}$ and sfermion mass
matrices as free parameters, without referring to specific 
supersymmetry breaking or flavour models. 
On the phenomenological side,
we will be interested in low-energy processes that 
only involve ordinary fermions and photons (and possibly
goldstinos) as external particles\footnote{If sfermions
are also allowed to be external, other processes
could be considered. The flavour changing decays 
$\ft_i \to f_j \pz$ are obvious examples.}.

The paper is organized as follows. In Section~2, we
embed the MSSM in an effective lagrangian in which
supersymmetry is linearly realized and spontaneously
broken by the auxiliary component vev of some chiral 
superfield $Z$, as in \cite{asp}. The couplings
of the goldstino superfield $Z$ to matter
and gauge superfields generate both mass and
interaction terms for the component fields.
In Section~3, we use those interactions to compute 
the rate of flavour changing radiative decays,
such as $\mu \to e \gamma$ and analogous ones. 
In particular, we compare the contributions with 
goldstino exchange with the conventional ones which 
do not involve the goldstino multiplet. 
In Section~4, we discuss the
generation of flavour changing effective operators 
with four external matter fermions, and 
discuss the phenomenological implications for
processes such as $\mu \to e e e$,
$K_L \to \mu^+ \mu^-$ and $K$-${\ov K}$
transitions. In Section~5, we address the 
issue of flavour violation in processes
involving two matter fermions and two
goldstinos as external states. 
In this case, we also consider a more general approach 
based on the non-linear realization of supersymmetry,
generalizing earlier results \cite{bfzj,cllw}.
We compare this approach with the linear one, also 
extending the latter to the case of mixed $F$-$D$ 
breaking.
Section~6 is devoted to summary and conclusions.  


\vspace{1 cm}

\noindent
{\large \bf 2. Supersymmetry breaking masses and goldstino couplings}
\vspace{0.3 cm}

The MSSM contains quark, lepton and Higgs supermultiplets interacting 
with the $SU(3)_c\times SU(2)_L\times U(1)_Y$ vector supermultiplets.
For our purposes, it will be sufficient to consider quark and 
charged lepton supermultiplets,  
coupled to the electromagnetic $U(1)_Q$ vector supermultiplet.
The matter supermultiplets in each charge sector will be generically 
denoted by $E_i=(\ft,f)_i$ (charge $Q_f$) and $E^c_i=(\fct,\fc)_i$ 
(charge $Q_{\fc}=-Q_f$), where $f=u,d,\ell$, and 
$i$ is a generation (i.e. flavour) index.
The $U(1)_Q$ vector supermultiplet contains photon ($A_{\mu}$) and 
photino ($\la$) fields. The associated lagrangian for the component
fields reads\footnote{We use two component spinor notation, with  
$\smu \! \equiv \! (1,\vec{\sigma})$, 
$\smub \! \equiv \!(1,-\vec{\sigma})$,
$\smn \! \equiv \! {1\over 4}(\smu\snub-\snu\smub)$, 
$\smnb \! \equiv \! {1\over 4}(\smub\snu-\snub\smu)$ and
$g_{\mu\nu} \! = \! {\rm diag}(+1,-1,-1,-1)$.
We recall that $f$ and $\bar{f}^c$ correspond to the left
and right components of the four component Dirac spinor $\Psi_f$,
whereas $\ft$ and $\fct^*$ correspond to the fields usually
denoted as $\fl$ and $\fr$.}
\bea
\label{lzero}
{\cl}_0 
& = & 
-{1\over 4} F_{\mu\nu} F^{\mu \nu} + i \lab \smub \dmd \la 
+ {1 \over 2} (M \la \la + {\rm h.c.})
+ \sum_{f=u,d,\ell} \left[
i \fb \smub D_{\mu} f + i \fcb \smub D_{\mu} \fc 
\right.
\nonumber\\ & & 
+ (D^{\mu} \ft)^* (D_{\mu} \ft)
+ (D^{\mu} \fct)^* (D_{\mu} \fct)
+ g_e \sqrt{2} Q_f (i \ft^* f \la - i \fct^* f^c \la  + {\rm h.c.})
\nonumber\\ & & 
\nonumber\\ & & 
\left.
- (\fc m f + {\rm h.c.})
- \left( \begin{array}{cc}\! \ft^*\! & \!\fct\! \end{array} \right)
\left( \begin{array}{cc} 
m^{\dagger}m + \mll & \mlr \\ \mrl & m m^{\dagger}  + \mrr   
\end{array} \right)
\left( \begin{array}{c} \ft \\ \fct^* \end{array} \right) 
\right]
+ \ldots
\eea
where $g_e$ is the electromagnetic gauge coupling, 
$D_{\mu}= \dmd + i g_e A_{\mu} Q $
and the dots denote sfermion self-interactions. 
Note that $m$, $\mll$, $\mrr$, $\mrl$, $\mlr=(\mrl)^{\dagger}$ 
are $3 \times 3$ matrices in each charged sector and should
be labelled by an index $f$. Both this index
and generation indices are understood for simplicity\footnote{
The inclusion of neutrinos requires
minor modifications. Consider the case in which only $\nu$ and $\tilde{\nu}$ 
are present in the low-energy theory. Then, wherever a sum $\sum_f$
appears or is understood, the neutrino contributions can be formally obtained 
by putting $f \!= \!f^c \!= \!\nu$, $\ft \!= \!\fct \!= \!\tilde{\nu}$, 
$\mll \!= \!\mrr$ and
multiplying by a factor $1/2$. For instance, the neutrino and sneutrino 
mass terms in eq.~(\ref{lzero}) would read as 
$-{1 \over 2}(\nu m \nu + {\rm h.c.})$ and 
$-\tilde{\nu}^* (m^{\dagger}m + \mll) \tilde{\nu}$ 
$-{1\over 2} (\tilde{\nu} \mrl\tilde{\nu}+{\rm h.c.})$.}.
If fermion and sfermion mass matrices are not diagonal in the same 
superfield basis, flavour changing effects arise through the 
gaugino-fermion-sfermion vertices. 
Notice that we do not assume that the matrix $\mrl$ 
is proportional to the matrix $m$. However, in order to simplify 
the power counting in the next sections, we will make the reasonable
assumption that both matrices have a common chiral suppression.
Thus, in the fermion mass basis, the diagonal entries of $\mrl$
are expected to be of order `supersymmetry breaking mass' 
$\times$ `appropriate fermion mass', and a further (model
dependent) suppression factor can be expected in the 
off-diagonal entries. 

Now we have to specify how the matter and gauge multiplets couple
to the goldstino, without relying on some specific fundamental
mechanism for supersymmetry breaking. This can be done in several ways. 
For instance, the interactions between one goldstino and a 
fermion-sfermion pair, which are model-independent, can 
easily be derived from supercurrent conservation \cite{fayet1}. 
For interactions involving more than one goldstino, other methods 
have to be used. Here we follow the approach of ref.~\cite{asp},
where the spontaneous breaking of supersymmetry is
described at an effective level. Therefore we will
consider the above lagrangian (\ref{lzero}) as part of an 
effective globally supersymmetric lagrangian in which the matter 
and gauge superfields are also coupled to some neutral chiral 
superfield $Z$. Supersymmetry, which is linearly realized, 
is assumed 
to be spontaneously broken by the auxiliary component of $Z$, 
through a non-vanishing expectation value $<Z|_{\th\th}>=-F$. 
The mass parameter $\sqrt{F}$ is the 
supersymmetry breaking scale.
The physical components in $Z$ are a Weyl fermion, namely the 
goldstino $\gol$, and a complex scalar $z$, called 
sgoldstino\footnote{We work in field coordinates
such that $<\!z\!>=0$, $<K_{\ov{z} z}>=1$ and the parameters $F$ 
and $M$ are real and positive.}. 
The effective couplings between the goldstino superfield $Z$ 
and matter or gauge superfields 
generate not only the supersymmetry breaking masses shown 
in $\cl_{0}$, through the above vev, but also 
closely related interactions, to be illustrated now.

Matter fermion masses, as well as the associated `supersymmetric'
sfermion masses, can be derived from superpotential terms
of the form $E^c m E$.
Consider now the supersymmetry breaking mass matrices in the sfermion
sector, i.e. $\mll$, $\mrr$ and $\mrl$. The $LL$ ($RR$) mass terms 
can be derived from  K\"ahler potential terms of the form $|Z|^2 E^* E$ 
($|Z|^2 E^c E^c\,^* $), suppressed by the square of some scale 
$\tilde\Lambda$, whereas the $RL$ mass terms 
can be derived from superpotential terms of the form
$Z E^c E$. All such terms contain arbitrary dimensionless flavour 
matrices. Since these matrices are in one-to-one 
correspondence with the matrices $\mll$, $\mrr$ and $\mrl$,
we will trade the former set of parameters 
(plus $\tilde\Lambda$) for
the latter set of physical parameters (including $\sqrt{F}$).
Those K\"ahler potential and superpotential terms generate
not only masses, but also
several interactions (of dimension 4 or higher) involving 
the physical components of the matter and goldstino  
superfields\footnote{
The effective supersymmetric lagrangian could contain other 
superfield interactions besides those considered here. Some of them 
can be eliminated in favour of the existing ones through 
field redefinitions. Other ones depend on additional 
arbitrary parameters, not directly related to the mass
spectrum. We will not discuss them, since we choose to focus 
on couplings that are related to the mass spectrum.}.
In particular, the following cubic interactions emerge:
\be
\label{cubic}
- {1 \over F} \sum_f 
\left( \begin{array}{cc} \ft^* & \fct \end{array} \right)
\left( \begin{array}{cc} 
\mll & \mlr \\ \mrl & \mrr   
\end{array} \right)
\left( \begin{array}{c} f\pz \\ \fcb \pzb \end{array} \right) 
+ {\rm h.c.} 
\ee
\be
- {1 \over F} \sum_f z f^c \mrl f  + {\rm h.c.} 
\ee
Among quartic interactions, the following ones will be 
relevant for our purposes:
\be
\label{quartic}
 - {1 \over 2 F^2} \left[ \pzb \smdb \pz + 
i( z^* \dmd z - z \dmd z^*) \right]
\sum_f (\fb \mll \smub f - \fc \mrr \smu \fcb) 
\ee
We stress again that the interactions of a goldstino with fermion-sfermion 
bilinears in (\ref{cubic}) are completely model independent, as
one can easily check by using supercurrent conservation.
As already mentioned, such interactions resemble those of a
neutral gaugino, with $\mt^2/F$ playing the role of $g$ \cite{fayet1}. 
Notice however that, in contrast to the case of gauginos, here the 
coupling already has a flavour structure. However, this is not a 
new independent structure: it is uniquely specified by the mass 
matrices, which also dictate the 
subsequent rotation to the physical bases. So the interactions 
of neutral gauginos and those of goldstinos are expected to 
have a similar impact on FCNC processes, once the parameter 
mapping is taken into account.
Nevertheless, the fact that the goldstino is the goldstone particle 
of spontaneously broken supersymmetry makes it quite special.
The peculiar low-energy properties of goldstinos will especially 
emerge in our final section.
 
As regards the gaugino mass term in $\cl_0$, one can 
effectively derive it from a superfield coupling of the form
$(\eta / \tilde\Lambda) Z \cal{W} \cal{W}$ 
(where $\cal{W}$ is the gauge superfield strength), i.e. from 
a linear term in the gauge kinetic function.
The associated interactions involving the physical components
of the vector and goldstino superfields can
be written in terms of $M$ and $F$ \cite{asp}:
\be  
- {M \over \sqrt{2} F} \pz \sigma^{\mu\nu} \la F_{\mu\nu}
- {M \over 4 F} z ( F_{\mu\nu} F^{\mu \nu}
+i F_{\mu\nu} \tilde{F}^{\mu \nu} ) +{\rm h.c.} + \ldots
\ee
where $\tilde{F}^{\mu \nu} ={1\over 2} \epsilon^{\mu\nu\rho\sigma}
F_{\rho\sigma}$ and the dots stand for terms which we will not need. 
Other superpotential and K\"ahler potential terms generate 
kinetic, mass and interaction terms in the $Z$ sector \cite{asp}:
\bea
& &  i \pzb \smub \dmd \pz + 
{1\over 2}(\dmu S \dmd S - m_S^2 S^2)
+{1\over 2}(\dmu P \dmd P - m_P^2 P^2)
\nonumber\\
& & - {1 \over 2 \sqrt{2} F} ( m_S^2 S \pz \pz - i m_P^2 P \pz \pz
+ {\rm h.c.} )
-{m_S^2 + m_P^2 \over 8 F^2}\pz \pz \, \pzb \, \pzb + \ldots
\eea
We have assumed for simplicity that the mass eigenstates in the 
sgoldstino sector coincide with the real and imaginary parts 
of $z=(S+iP)/\sqrt{2}$. 
We will also assume that sgoldstino masses are not much lighter 
than squark and slepton masses, as suggested by naturalness 
considerations \cite{bfzff}. If this assumption is relaxed 
and sgoldstinos are allowed to be very light, enhancement 
effects can appear in several processes 
(see e.g. \cite{lights,bpz}),
including some FCNC processes to be discussed below. 
However, in such cases $\sqrt{F}$ is typically forced to be 
substantially larger than the electroweak scale, which is not 
the scenario we would like to study\footnote{Collider signals 
of massive sgoldstinos in the case of low $\sqrt{F}$ have been
recently analysed in \cite{prz}.}.


\vspace{1 cm}

\noindent
{\large \bf 3. Flavour changing radiative decays}
\vspace{0.3 cm}

In this section we will discuss flavour changing radiative decays.
For definiteness, we focus on the decay $\mu^- \to e^- \gamma$, which
violates individual lepton flavours. 
The effective operator responsible for such a decay can be
parametrized as 
\be
\cl_{eff} = {i g_e \over 16 \pi^2} m_{\mu} 
( C_R {\ov e} \smnb {\ov\mu}^c + C_L e^c \smn \mu ) F_{\mu\nu}
\ee
This leads to the branching ratio
\be
BR(\mu\to e \gamma) =  {3 \alpha \over 16 \pi G_F ^2} (|C_R|^2+|C_L|^2)
\, BR(\mu \to e \nu_{\mu} \ov\nu_e) 
\ee
where $BR(\mu \to e \nu_{\mu} \ov\nu_e)\simeq 1$. 

The interactions involving goldstinos and sgoldstinos described in
the previous section generate several one-loop contributions
to $C_R$ and $C_L$, through the diagrams schematically shown 
in Fig.~1.
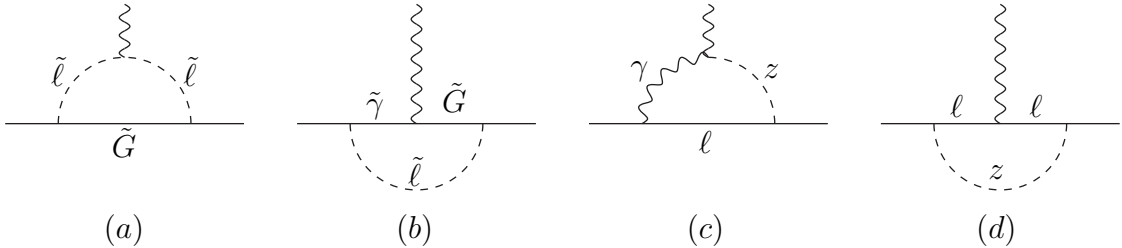
\begin{figure}[htb]
\begin{center}
\begin{picture}(440,100)(0,0)
\Line(10,50)(100,50)
\DashCArc(55,50)(25,0,180){3}
\Photon(55,75)(55,95){2}{3}
\Text(55,43)[]{$\gol$}
\Text(30,70)[]{$\lt$}
\Text(80,70)[]{$\lt$}
\Text(55,10)[]{$(a)$}
\Line(120,50)(210,50)
\DashCArc(165,50)(25,180,0){3}
\Photon(165,50)(165,95){2}{6}
\Text(150,57)[]{$\tilde{\ga}$}
\Text(180,60)[]{$\gol$}
\Text(165,32)[]{$\lt$}
\Text(165,10)[]{$(b)$}
\Line(230,50)(320,50)
\DashCArc(275,50)(25,0,90){3}
\PhotonArc(275,50)(25,90,180){2}{5}
\Photon(275,75)(275,95){2}{3}
\Text(275,43)[]{$\ell$}
\Text(250,70)[]{$\ga$}
\Text(300,70)[]{$z$}
\Text(275,10)[]{$(c)$}
\Line(340,50)(430,50)
\DashCArc(385,50)(25,180,0){3}
\Photon(385,50)(385,95){2}{6}
\Text(370,57)[]{$\ell$}
\Text(400,57)[]{$\ell$}
\Text(385,32)[]{$z$}
\Text(385,10)[]{$(d)$}
\end{picture}
\end{center}
\vspace{-0.5cm}
\caption{\em Diagrams with goldstino or sgoldstino
exchange contributing to $\mu \to e \gamma$.} 
\label{fig1}
\end{figure}
We disregard type (d) diagrams, since they are quadratic in $\mlr$
and we are only working at first order in the muon mass.
From the other diagrams, we obtain:
\bea
C_R^{(G)} \!\! & \!\! =\!\! &  \!\!
{1 \over 2 F^2} \! \left[ -{1 \over 6} \mll + 
\left(  { \mmt^2 M^2 \over \mmt^2 - M^2 }
\! \left(1- {M^2 \over \mmt^2 - M^2 } \log{\mmt^2 \over M^2}
\right) \! \right)_{LL} 
\!\! + \! \left( \log{m_P^2 \over m_S^2}\right) 
\! {M \over m_{\mu}} \mlr
\right]_{e \mu}
\\
C_L^{(G)} \!\! & \!\!  =\!\!  & \!\! 
{1 \over 2 F^2} \! \left[ -{1 \over 6} \mrr + 
\left( { \mmt^2 M^2 \over \mmt^2 - M^2 }
\! \left(1- {M^2 \over \mmt^2 - M^2 } \log{\mmt^2 \over M^2}
\right) \! \right)_{RR} 
\!\! + \! \left( \log{m_P^2 \over m_S^2}\right) 
\! {M \over m_{\mu}} \mrl
\right]_{e \mu}
\eea
where $\mmt^2$ stands for the $6 \times 6$ slepton mass matrix,
whose $3\times 3$ blocks are $\mll$, $\mrr$, $\mlr$, $\mrl$.
These results generalize to flavour-changing transitions those 
obtained for diagonal magnetic moments in the absence of 
flavour mixing \cite{magn,bpz}.
The above expressions hold in the superfield basis in which leptons 
are mass eigenstates. The computation can be performed by 
using matrix vertices and propagators in that basis.
Alternatively, one can diagonalize the slepton mass matrix
as well and move back to the other basis in the end.
The three terms in each expression originate from diagrams
of type (a), type (b) and type (c), respectively.
Notice that type (a) contributions are simply proportional 
to the ${e \mu}$ element of the matrices $\mll$ and $\mrr$.
The latter result holds for arbitrary $\mmt^2$: it does not 
rely on any assumption on the size of the off-diagonal entries 
of $\mll$ and $\mrr$, or even on the assumption that the 
entries of $\mrl$ are linear in lepton masses. 
The matrix structure of vertices and propagators 
just combine in the proper way, and the final result
turns out to coincide with what we would have obtained 
in the simple mass insertion approximation\footnote{
Incidentally, we remark that even the mass insertion method
has new features in the present context. For type (a) 
diagrams, for instance, the flavour violating factor $(\mll)_{e\mu}$ 
(or $(\mrr)_{e\mu}$) can be inserted in either a slepton 
{\em propagator} or a lepton-slepton-goldstino {\em vertex}. 
Moreover, the vertex contributions are twice as large as the 
propagator contributions and have opposite sign.}. 
Type (c) contributions are proportional to the $e \mu$ elements 
of $\mlr$ and $\mrl$, but here the reason is more obvious. 
On the other hand, type (b) contributions are expressed 
through a non-trivial function of the matrix $\mmt^2$. 
In this case, in order to obtain an approximate expression,
we can expand $\mmt^2$ around the diagonal\footnote{
For simplicity, we will consider the diagonal entries of 
$\mmt^2$ to have a common value $\mt^2$. The generalization 
is straightforward.}
and work to first order in the flavour changing elements 
of $\mmt^2$, now assumed to be small.
This corresponds to the mass-insertion approximation
and allows us to cast type (b) contributions in a form similar
to the other ones. Under this approximation, we can rewrite
$C_R^{(G)}$ and $C_L^{(G)}$ as
\bea
C_R^{(G)} & = & 
{1 \over F^2} \left[ H_1 \left({M^2\over \mt^2}\right)(\mll)_{e\mu} + 
{1\over 2} \left( \log{m_P^2 \over m_S^2}\right)
{M (\mlr)_{e\mu} \over m_{\mu}} \right] 
\\
C_L^{(G)} & = & 
{1 \over F^2} \left[H_1\! \left({M^2\over \mt^2}\right)(\mrr)_{e\mu} + 
{1\over 2} \left(\log{m_P^2 \over m_S^2}\right)
{M (\mrl)_{e\mu} \over m_{\mu}} \right] 
\eea
where
\be
H_1 (x) = { -1 + 3 x - 15 x^2 + 13 x^3 - 6 x^2 (1+x)\log x
\over 12 (1-x)^3 }
\ee
The function $H_1(x)$ is negative for $x < 1$ and positive
for $x > 1$. Notice that $H_1(1)=0$: when $\mt^2 = M^2$,
type (a) and type (b) contributions cancel each other,
to linear order in the flavour changing masses. 
 
We would like to compare the above contributions, which we
will simply call `goldstino contributions', to the more
conventional non-goldstino contributions. In the full MSSM, 
the latter ones arise from both charged and neutral 
interactions. The reference lagrangian $\cl_0$
only gives neutral contributions, from type (a) diagrams 
in which the goldstino is replaced by a photino. 
In contrast to the goldstino, however, the photino propagator 
can either conserve or flip chirality, so several contributions 
arise. In the mass-insertion approximation, we find 
(in agreement with \cite{bor,gab}):
\bea
C_R^{(0)} \!\! &  \!\! =  \!\! &  \!\!{2 g_e^2 \over \mt^4}
\! \left[\! \left(\! H_2\! \left({M^2\over \mt^2}\right) 
+ H_3\! \left({M^2\over \mt^2}\right) \! 
{M (\mlr)_{\mu \mu} \over \mt^2 \, m_{\mu}} \right) (\mll)_{e\mu} 
- H_4\! \left({M^2 \over \mt^2}\right) \!
{M (\mlr)_{e\mu} \over m_{\mu}} \right] 
\\
C_L^{(0)}\!\! &\!\! =\!\! &\!\!  {2 g_e^2 \over \mt^4}
\!\left[ \!\left(\! H_2\! \left({M^2\over \mt^2}\right) 
+ H_3\! \left({M^2\over \mt^2}\right)\! 
{M (\mrl)_{\mu \mu} \over \mt^2 \, m_{\mu}} \right) (\mrr)_{e\mu} 
- H_4\! \left({M^2 \over \mt^2}\right) \!
{M (\mrl)_{e\mu} \over m_{\mu}} \right] 
\eea
where the loop functions $H_2(x), H_3(x), H_4(x)$ are
\bea
H_2(x)& =& {1- 9 x - 9 x^2 + 17 x^3 - 6 x^2 ( 3 + x) \log x
\over 12 (1-x)^5 }
\\
H_3(x)& = & 
{1 + 9 x - 9 x^2 - x^3 + 6 x ( 1 + x) \log x
\over 2 (1-x)^5 }
\\
H_4(x)& = & {1 + 4 x  - 5 x^2 + 2 x (2 + x) \log x
\over 2(1-x)^4 }
\eea

Goldstino and non-goldstino contributions exhibit
a similar structure. They are proportional to the flavour 
changing ($e\mu$) elements of the slepton mass matrix, 
depend on dimensionless functions of superpartner masses
and are suppressed by the fourth power of some scale.
This scale is the supersymmetry breaking scale
for the goldstino contributions and a supersymmetry
breaking mass (e.g. the average slepton mass $\mt$) 
for the non-goldstino contributions.
If $\sqrt{F}$ is much larger than the supersymmetry
breaking masses, the goldstino contributions are negligible 
in comparison to the non-goldstino ones. On the other
hand, if $\sqrt{F}$ and the supersymmetry breaking masses
have a similar size, then goldstino and non-goldstino
diagrams give similar contributions to $\mu \to e \gamma$.
It is interesting to make a more quantitative 
comparison. For definiteness, we neglect LR terms (both
flavour conserving and flavour changing ones) and focus 
on the contributions proportional to $(\mll)_{e\mu}$,
i.e. we consider
\be
C^{(G)}_R = {1 \over F^2} 
H_1 \! \left({M^2 \over\mt^2}\right) (\mll)_{e\mu}
\;\; ,
\;\;\;\;\;\;\; 
C^{(0)}_R = {2 g_e^2 \over \mt^4} 
H_2 \! \left({M^2 \over\mt^2}\right) (\mll)_{e\mu}
\ee
We recall that $H_1(x)$ can have either sign, whereas
$H_2(x)$ is positive. 
To measure the relative importance of the goldstino contributions 
versus the non-goldstino ones, we introduce the ratio
\be
R = {C^{(G)}_R \over C^{(0)}_R} = {\mt^4 \over 2 g_e^2 F^2} 
\dfrac{H_1 \! \left( \frac{M^2}{\mt^2} \right)} 
{H_2 \! \left( \frac{M^2}{\mt^2} \right)}
\ee
which does not depend on $(\mll)_{e\mu}$. 
In the limit $\mt \gg M$, for instance, that ratio
is $R = - \mt^4/(2 g_e^2 F^2)$, which becomes $-1$ when 
$\mt \simeq 0.65 \, \sqrt{F}$. Contours of $R$ in the
$(\mt/\sqrt{F},M/\sqrt{F})$ plane are shown in 
Fig.~2. 
Goldstino contributions are smaller (larger) than
the other ones in the region with $|R| < 1$ ($|R| > 1$).
We can also combine the two classes of contributions and 
study $BR(\mu \to e \gamma)$ as a function of $\mt$, $M$, 
$\sqrt{F}$ and $(\delta_{LL})_{e\mu} \equiv (\mll)_{e\mu}/\mt^2$.
\begin{figure}[p]
\vskip -0.7cm
\centerline{\protect\hbox{
\epsfig{file=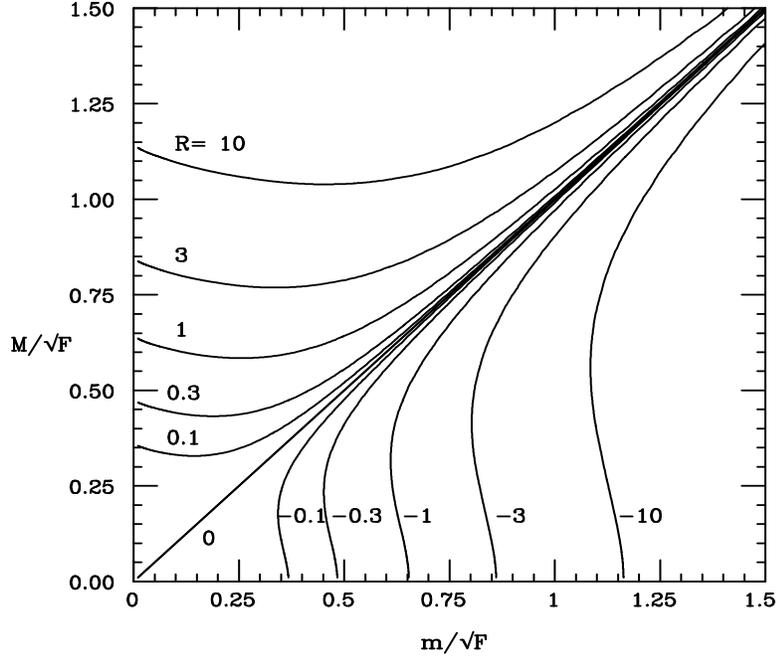,height=16.cm,width= 10.9cm,angle=90}}}
\vskip -1.3cm
\caption{\em The ratio of goldstino versus non-goldstino 
contributions to the $\mu \to e \gamma$ amplitude in the 
$(\mt/\sqrt{F},M/\sqrt{F})$ plane.}
\label{f2}
\end{figure}
\begin{figure}[p]
\vskip -0.7cm
\centerline{\protect\hbox{
\epsfig{file=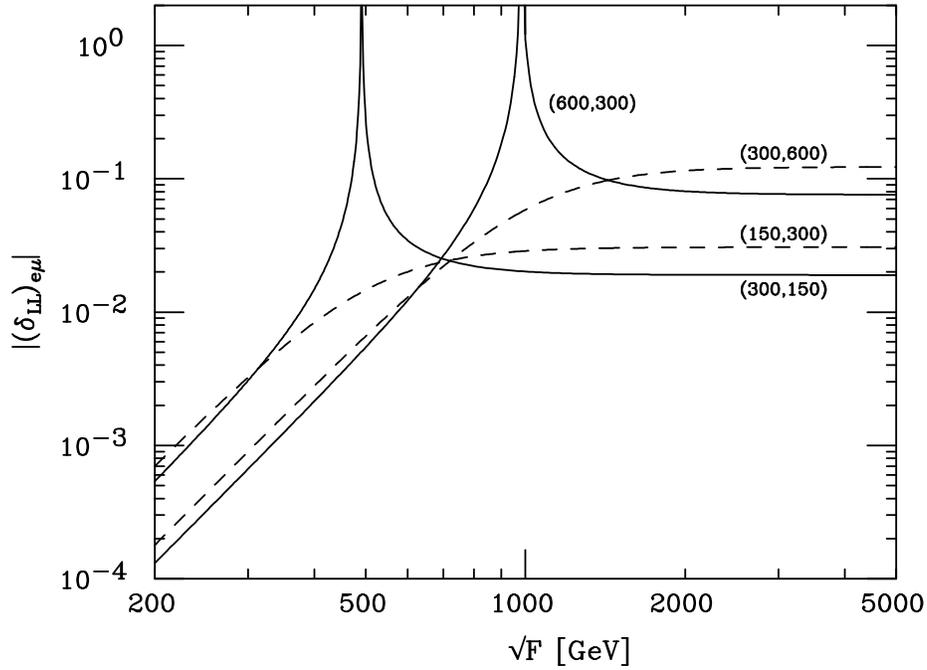,
width= 10.9cm,angle=90}}}
\vskip -1.cm
\caption{\em 
Lines of constant $BR(\mu \to e \gamma)=1.2 \times 10^{-11}$
in the $(\sqrt{F},|(\delta_{LL})_{e\mu}|)$ plane, for 
different choices of $(\mt,M) \, [ {\rm GeV}] $.}
\vskip -0.5cm
\label{f3}
\end{figure}
In Fig.~3 we have fixed
some representative values of $(\mt,M)$ and shown
the lines in the $(\sqrt{F},|(\delta_{LL})_{e\mu}|)$ 
plane along which $BR(\mu \to e \gamma)$ saturates 
the present experimental bound, which is $1.2 \times 10^{-11}$
\cite{mega}. 
In other words, the lines give the upper bound 
on $|(\delta_{LL})_{e\mu}|$ as a function of $\sqrt{F}$.
When $\sqrt{F}$ is large such bounds are determined by 
the (conventional) non-goldstino contributions.
When $\sqrt{F}$ decreases, the latter contributions
start to interfere with the goldstino ones: for $\mt >M$
the interference is destructive, the bound on 
$|(\delta_{LL})_{e\mu}|$ tends to disappear and the curves
exhibit a peak, whereas for $\mt < M$ the interference 
is constructive and the curves show a knee. 
For even smaller values of $\sqrt{F}$,
i.e. to the left of the transition region, the 
goldstino contributions dominate and the bounds on 
$|(\delta_{LL})_{e\mu}|$ become stronger than the 
conventional ones.  
In the limit $\mt \gg M$, for instance, the bound from 
goldstino contributions alone can be written as\footnote{
Incidentally, we recall that perturbativity considerations
require that the ratio $\mt/\sqrt{F}$ be smaller
than $2 \div 3$ \cite{asp,bfzff}: this should be
understood everywhere.
We also notice that the inequality (\ref{megbound})
can be equivalently written as 
$(\mll)_{e\mu} \simlt (13 {\rm GeV})^2 (\sqrt{F}/
 300 {\rm GeV})^4$, i.e. $\mt^2$ drops out. 
This example shows that, in the case of goldstino
contributions, the parametrization in terms of 
$(\delta_{LL})_{e\mu}$ and $\mt^2$ may be
redundant. We adopt it to allow for an easier
comparison with the literature.}  
\be
\label{megbound}
|(\delta_{LL})_{e\mu}| \simlt 2 \times 10^{-3}
\left( { \sqrt{F} \over \mt} \right)^4 
\left( {\mt \over 300 \, {\rm GeV} } \right)^2
\ee
An identical discussion applies
to the contributions that depend on $(\delta_{RR})_{e\mu}$.

It is straightforward to translate the above discussion
to the decays $\tau \to e \gamma$ and $\tau \to \mu \gamma$, 
whose branching ratios are experimentally bounded
by $ 2.7 \times 10^{-6}$ \cite{pdg} and 
$1.1 \times 10^{-6}$ \cite{cleo}, respectively. 
Notice that Fig.~2 applies to these cases 
as well, whereas in Fig.~3 only the scale of the vertical 
axis has to be changed: $|(\delta_{LL})_{e\mu}|$ 
has to be replaced by either 
$10^{-3} |(\delta_{LL})_{e\tau}|$ 
or $1.4 \times 10^{-3} |(\delta_{LL})_{\mu\tau}|$. Therefore
the qualitative description remains the same as before,
although the constraints on $|(\delta_{LL})_{e\tau}|$ 
and $|(\delta_{LL})_{\mu\tau}|$ are of course
much weaker. 

The above discussion can also be extended to flavour changing
transitions in the quark sector, such as $b \to s \gamma$. 
This decay is potentially sensitive to the $bs$ entries of 
the down squark mass matrix, through both non-goldstino and 
goldstino contributions. The former ones are mainly due to 
squark-gluino (rather than squark-photino) exchange 
\cite{bbm,gab}.
The latter ones become comparable to those when $\mt^2/F \sim g_s$, 
where $g_s$ is the strong coupling constant and $\mt$ is an 
average squark mass. However, neither contribution gives 
significant constraints. 

  
\vspace{1 cm}

\noindent
{\large\bf 4. Flavour changing processes with four matter fermions}
\vspace{0.3 cm}

We now discuss the FCNC processes that involve four matter 
fermions as external states.
In both the SM and the MSSM, the leading perturbative 
contributions to such processes generically arise at 
one-loop level, and are finite. 
If goldstino and sgoldstino couplings are also taken 
into account, additional contributions arise. 

Some contributions arise already at tree level. Indeed, when two
$z f \fc$ vertices are joined by a sgoldstino propagator,
effective four fermion interactions are 
generated:
\be
\label{fourrl}
\cl_{eff} = {1 \over 4 F^2} \left[ {1 \over m_S^2} 
(\sum_{f} f^c \mrl f + {\rm h.c.})^2 + {1 \over m_P^2} 
(i \sum_{f} f^c \mrl f + {\rm h.c.})^2 \right]
\ee
However, owing to the assumed chiral suppression of
$\mrl$ (and the assumed size of sgoldstino masses), 
the coefficients of these four fermion operators are at most 
$\co (m_f^2/F^2) \simlt \co (m_f^2 G_F^2)$, i.e. they are
automatically suppressed by fermion masses\footnote{For a
similar reason, other interactions due to tree-level 
sgoldstino exchange are also suppressed. 
For instance, by connecting a $z f \fc$
vertex with a $z \gamma \gamma$ vertex, we obtain
the effective (two-fermion)-(two-photon) coupling 
${M \over 4 F^2} (\sum_{f} f^c \mrl f)
({1 \over m_S^2} F_{\mu\nu} F^{\mu \nu}
- {i \over m_P^2}  F_{\mu\nu} \tilde{F}^{\mu \nu})
+ {\rm h.c.}$,
which could contribute e.g. to the decay $\mu \to e \ga \ga$.}.
On the other hand, if the above mentioned assumptions are
relaxed, enhancement effects can appear: in this case,
the required suppression should be provided by 
large values of $\sqrt{F}$ and/or intrinsically small
flavour violation in $\mrl$. 

We would like to discuss contributions to four fermion
processes that are not chirally suppressed, 
so we neglect $\mrl$ and move to one-loop level. 
In analogy to the SM or the MSSM, several one-loop diagrams 
contribute, both 1PI (e.g. boxes) and 1PR (e.g. penguins).
Some 1PI diagrams are skecthed in Figure~4. Diagram (a)
is a non-goldstino MSSM diagram, which we show for comparison:
it is a typical gaugino-sfermion box. By replacing 
gaugino-fermion-sfermion vertices with  
goldstino-fermion-sfermion vertices, we obtain
box diagrams like (b) and (c). All these boxes give finite 
contributions\footnote{For instance, the coefficients
of $\Delta F=1$ four fermion operators induced by
(a), (b) and (c) scale as 
$(g^4 / \mt^4) \mt^2_{ij}$, $(g^2 / F^2) \mt^2_{ij}$ 
and $(\mt^4 / F^4) \mt^2_{ij}$, respectively,
where $\mt^2_{ij}$ is the appropriate flavour changing
sfermion mass.}. In our supersymmetric effective lagrangian, 
however, goldstinos (and sgoldstinos) also couple to matter 
through non-renormalizable couplings. For example, the theory 
contains quartic interactions of dimension six, like 
those in eq.~(\ref{quartic}), whose coefficients are
again determined by sfermion masses and $\sqrt{F}$.
These interactions cannot be simply dropped, since
they play a crucial role in the low-energy cancellations 
that take place in diagrams with external goldstinos 
(see next section). Such vertices can be used to 
build new (non box) 1PI diagrams, like (d), (e), (f) 
in Figure~4.
These diagrams are not finite: the dependence on
the cutoff scale is logarithmic for diagram (d)
and quadratic for diagrams (e) and (f). 
\begin{figure}[htb]
\begin{center}
\begin{picture}(360,150)(0,0)
\Line(20,100)(100,100)
\Line(20,150)(100,150)
\DashLine(35,100)(35,150){4}
\DashLine(85,100)(85,150){4}
\Text(20,100)[r]{$f \,$}
\Text(20,150)[r]{$f \,$}
\Text(35,125)[r]{$\ft \,$}
\Text(60,110)[]{$\lambda$}
\Text(60,143)[]{$\lambda$}
\Text(100,100)[l]{$\, f$}
\Text(100,150)[l]{$\, f$}
\Text(85,125)[l]{$\, \ft$}
\Text(60,85)[]{$(a)$}
\Line(140,100)(220,100)
\Line(140,150)(220,150)
\DashLine(155,100)(155,150){4}
\DashLine(205,100)(205,150){4}
\Text(140,100)[r]{$f \,$}
\Text(140,150)[r]{$f \,$}
\Text(155,125)[r]{$\ft \,$}
\Text(180,110)[]{$\gol$}
\Text(180,143)[]{$\lambda$}
\Text(220,100)[l]{$\, f$}
\Text(220,150)[l]{$\, f$}
\Text(205,125)[l]{$\, \ft$}
\Text(180,85)[]{$(b)$}
\Line(260,100)(340,100)
\Line(260,150)(340,150)
\DashLine(275,100)(275,150){4}
\DashLine(325,100)(325,150){4}
\Text(260,100)[r]{$f \,$}
\Text(260,150)[r]{$f \,$}
\Text(275,125)[r]{$\ft \,$}
\Text(300,110)[]{$\gol$}
\Text(300,143)[]{$\gol$}
\Text(340,100)[l]{$\, f$}
\Text(340,150)[l]{$\, f$}
\Text(325,125)[l]{$\, \ft$}
\Text(300,85)[]{$(c)$}
\Line(20,0)(35,0)
\Line(20,50)(35,50)
\DashLine(35,0)(35,50){4}
\Line(35,0)(85,25)
\Line(35,50)(85,25)
\Line(85,25)(100,10)
\Line(85,25)(100,40)
\Text(20,0)[r]{$f \,$}
\Text(20,50)[r]{$f \,$}
\Text(35,25)[r]{$\ft \,$}
\Text(60,5)[]{$\gol$}
\Text(60,50)[]{$\gol$}
\Text(100,10)[l]{$\, f$}
\Text(100,40)[l]{$\, f$}
\Text(60,-15)[]{$(d)$}
\CArc(180,25)(25,0,360)
\Line(155,25)(140,10)
\Line(155,25)(140,40)
\Line(205,25)(220,10)
\Line(205,25)(220,40)
\Text(140,10)[r]{$f \,$}
\Text(140,40)[r]{$f \,$}
\Text(180,10)[]{$\gol$}
\Text(180,40)[]{$\gol$}
\Text(220,10)[l]{$\, f$}
\Text(220,40)[l]{$\, f$}
\Text(180,-15)[]{$(e)$}
\DashCArc(300,25)(25,0,360){4}
\Line(275,25)(260,10)
\Line(275,25)(260,40)
\Line(325,25)(340,10)
\Line(325,25)(340,40)
\Text(260,10)[r]{$f \,$}
\Text(260,40)[r]{$f \,$}
\Text(300,7)[]{$z$}
\Text(300,43)[]{$z$}
\Text(340,10)[l]{$\, f$}
\Text(340,40)[l]{$\, f$}
\Text(300,-15)[]{$(f)$}
\end{picture}
\end{center}
\vspace{0.5cm}
\caption{\em Examples of one-particle-irreducible contributions
to flavour changing four fermion operators. The symbols
$f$ and $\ft$ generically denote matter fermions and sfermions.
Flavour changes can occur in sfermion propagators, cubic 
vertices or quartic vertices.} 
\label{fig4}
\end{figure}
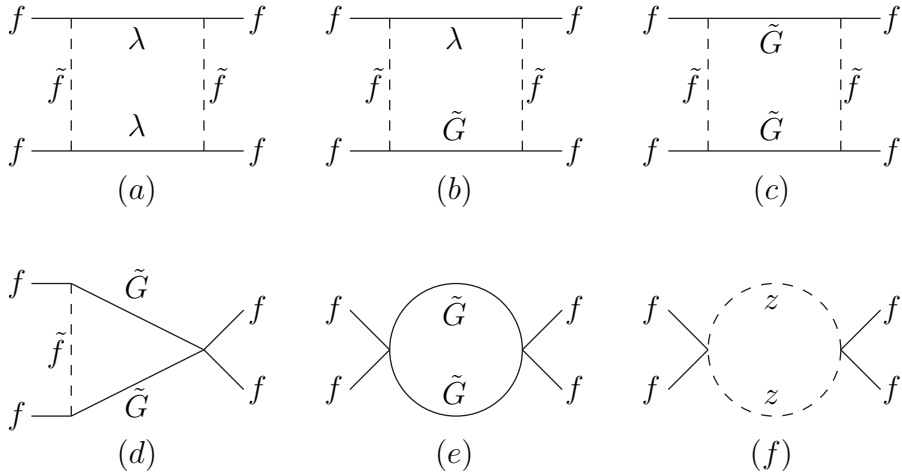
Thus, in contrast to what happens in the SM or the MSSM, 
flavour changing four fermion interactions receive both finite and
divergent contributions in the effective theory considered here. 
In principle, such divergences could be reabsorbed by introducing
other terms in the supersymmetric effective lagrangian, 
such as (for instance) 
K\"ahler potential terms quartic in the matter superfields.
These new terms would not only act as counterterms, but also
generate new contributions. 
Strictly speaking, all this means that the coefficients of 
four fermion interactions cannot be predicted in terms of the
parameters already introduced\footnote{The situation was slightly 
different in the computation of $\mu \to e \gamma$ presented in 
the previous section. In that case the operator was different, 
and our choice to only focus on couplings related to the spectrum 
led us to obtain a finite result.
If we had included other terms in the K\"ahler potential
or in the gauge kinetic function, however, we would also 
have obtained logarithmically divergent contributions,
associated to a (supersymmetric) higher derivative term. 
For a more detailed discussion in the context of
the flavour conserving magnetic moments, see \cite{bpz}.}. 
However, we can also adopt the milder point of view
that the one-loop diagrams generated by the interactions originally 
introduced give naturalness estimates of the coefficients of 
four-fermion interactions, once the cutoff scale is 
specified\footnote{For a similar discussion about this and
the previous points in the context of flavour conserving 
interactions, see \cite{bfzff}.}.
We continue the discussion in this spirit and focus on the
contributions generated by diagrams (e) and (f) in Fig.~4. 
By retaining only the quadratic dependence on the
cutoff scale $\Lambda$, we obtain:
\be
\label{fourfer}
\cl_{eff} = - {1 \over 64 \pi^2 }{\lsq \over F^4}
\left[\sum_f (\fb \mll  \smub f - \fc \mrr \smu \fcb)\right]
\left[\sum_{f'} (\fb' \mll  \smdb f' - 
f^{\prime c} \mrr \smd \ov{f}^{\prime c} )\right]
\ee
These four fermion terms can alternatively be extracted 
from the general formula $\Delta K = \dfrac{\lsq}{16 \pi^2} 
\log \det K_{\ov{m} n}$, which summarizes the quadratically 
divergent contribution of chiral supermultiplets to the 
K\"ahler potential \cite{grk}.

We will use the effective lagrangian (\ref{fourfer}) 
to estimate the effect of flavour changing goldstino 
(or sgoldstino) interactions on $\Delta F=1$ processes 
such as $\mu \to eee$, $K^0 \to \ell \ov{\ell}$,
$K \to \pi \ell \ov{\ell}$, $\ldots$, or $\Delta F=2$
processes such as $K$-$\ov{K}$ transitions.
As far as sfermion mass matrices are concerned,
we will assume for simplicity that, in the fermion mass
basis, the flavour diagonal entries in the $LL$ and $RR$ 
blocks have a common value $\mt^2$. 
We stress again that our results below should be regarded as
indicative, not only because we are focusing on a specific class 
of contributions, but also because the quadratic 
sensitivity on the cutoff scale $\Lambda$
introduces a further uncertainty. Indeed, in the
absence of information on the underlying theory, $\Lambda$ 
could either be a scale just above $\mt$ or take larger values,
up to the scale where unitarity breaks down. Hence   
$\Lambda$ is expected to lie somewhere 
in the range $[\Lambda_{min},\Lambda_{max}]$, 
where $\Lambda_{min}^2 \sim \mt^2$ and $\Lambda_{max}^2 \sim 
16 \pi F^2/\mt^2 \sim 16 \pi \tilde\Lambda^2$ 
\cite{asp,bfzff}. In the examples below, we will specialize
our formulae to the two extreme values of $\Lambda$.
We note that, for the conservative choice 
$\Lambda \sim \mt$, quadratically divergent contributions 
have similar parameter dependence and size as the other 
contributions (i.e. logarithmically divergent and finite ones), 
so in this case the former ones can also be interpreted as
`representatives' of the latter ones.

Consider for instance the terms in (\ref{fourfer}) 
that contribute to the lepton flavour violating 
process $\mu^- \to e^- e^+ e^-$:
\be
\cl_{eff} = - {1 \over 32 \pi^2 }{\mt^2 \lsq \over F^4}
( \ov{e} \smub e - e^c \smu \ov{e}^c)
\left[(\mll)_{e\mu} \ov{e}\smdb \mu 
- (\mrr)_{\mu e} e^c \smd \ov\mu^c\right]
\ee
If we focus on the part proportional to $(\mll)_{e\mu}$ and 
neglect any other contribution to this process, we obtain:
\be
BR(\mu^- \to e^- e^+ e^-) \simeq
\left[ { \sqrt{6} \over 128 \pi^2}  
{ \mt^4 \lsq \over G_F F^4}|(\delta_{LL})_{e\mu}| \right]^2
\ee
where $(\delta_{LL})_{e\mu}=(\mll)_{e\mu}/ \mt^2$.
Comparing the above expression with the experimental upper 
bound $10^{-12}$ \cite{pdg} gives constraints on the parameters
$(\delta_{LL})_{e\mu}$, $\mt$ and $\sqrt{F}$, for a
given $\Lambda$. For the two extreme choices of $\Lambda$, 
we obtain
\bea
\label{meeebound}
& & |(\delta_{LL})_{e\mu}| \simlt 5 \times 10^{-4}
\left( { \sqrt{F} \over \mt} \right)^8 
\left( {\mt \over 300 \, {\rm GeV} } \right)^2
\, ,  \;\;\;\;\;\; (\Lambda = \Lambda_{min})
\\
& & |(\delta_{LL})_{e\mu}| \simlt  10^{-5}
\left({ \sqrt{F} \over \mt} \right)^4
\left( {\mt \over 300 \, {\rm GeV} } \right)^2
\, ,  \;\;\;\;\;\;\;\;\;\;\;\; (\Lambda = \Lambda_{max})
\eea
If we take the smallest value of $\Lambda$, 
the constraints on $(\delta_{LL})_{e\mu}$ are comparable 
or stronger than those obtained in the previous section 
from the analysis of goldstino contributions to 
$\mu \to e \gamma$. This can be seen, for instance,
by comparing eq.~(\ref{meeebound}) with eq.~(\ref{megbound})
(also notice the different dependence on $\sqrt{F}/\mt$). 
If we take the largest possible value of $\Lambda$, the 
constraints on $(\delta_{LL})_{e\mu}$ are even stronger, 
for fixed values of the other parameters.
We recall that for large $\sqrt{F}$, i.e. when the 
non-goldstino contributions are the dominant ones, 
the process $\mu \to e \gamma$ is more sensitive to 
$(\delta_{LL})_{e\mu}$ as compared to $\mu \to eee$ \cite{bor}. 
Here we have found that when $\sqrt{F}$ approaches $\mt$, 
i.e. when goldstino contributions become important, 
the situation may be reversed. 
Similar considerations apply to $(\delta_{RR})_{e\mu}$, 
of course.

The lepton flavour violating decay $\pi^0 \to \mu e$ is 
another process that is sensitive to $(\mll)_{e\mu}$ and 
$(\mrr)_{e\mu}$. This decay receives contributions from terms 
in (\ref{fourfer}) which couple a muon-electron current to 
up or down quark currents. The latter currents couple to the pion
provided the up and down squark masses are non-degenerate.
However, owing to the rapidity of the dominant decay 
$\pi^0 \to \gamma \gamma$, no significant constraints are 
obtained on $(\delta_{LL})_{e\mu}$ and $(\delta_{RR})_{e\mu}$. 
Another process potentially sensitive to the latter quantities
is the $\mu \to e$ conversion on nuclei.
The effective interactions 
in (\ref{fourfer}) also contribute to other lepton flavour
violating processes, such as $\tau$ decays into either 
three charged leptons or a charged lepton and a $\pi^0$. 
These processes are sensitive to flavour changing 
slepton masses involving the third generation, 
but no strong constraints are obtained.

In the final part of this section, we consider some examples
of flavour violation in the quark sector induced by
the four-fermion terms in (\ref{fourfer}).  
In particular, we focus on two processes that are sensitive 
to the $sd$ entries of the squark mass matrix,
i.e. the decay $K_L \to \mu^+ \mu^-$ ($\Delta S=1$) and 
$K$-$\ov{K}$ transitions  ($\Delta S=2$).
Since we are dealing with order-of-magnitude estimates,
we neglect $QCD$ corrections and use the vacuum insertion
approximation.
The terms in (\ref{fourfer}) that contribute to the 
decay $K_L \to \mu^+ \mu^-$ are\footnote{These terms
or analogous ones (with the muon replaced by another
lepton or a quark) also contribute to other 
$\Delta S=1$ decays, such as $K \to \pi \ell \ov\ell$
or $K \to \pi\pi$ (hence $\epsilon'$). 
}
\be
\cl_{eff} = - {1 \over 32 \pi^2 }{\mt^2 \lsq \over F^4}
( \ov{\mu} \smub \mu - \mu^c \smu \ov\mu^c )
\left[(\mll)_{d s} \ov{d}\smdb s 
- (\mrr)_{d s} d^c \smd \ov{s}^c  + {\rm h.c.}\right]
\ee
If we only consider the part proportional to 
$(\mll)_{d s}$, we obtain:
\be
BR(K_L \to \mu^+ \mu^-) \simeq 
\xi \cdot \left[ { \sqrt{2} \over 32 \pi^2}  
{ \mt^4 \lsq \over \sin\theta_c G_F F^4}
{\rm Re}(\delta_{LL})_{ds} \right]^2
\ee
where $(\delta_{LL})_{d s}=(\mll)_{ds}/ \mt^2$, 
$\xi \equiv BR(K^+ \to \mu^+ \nu_{\mu}) \tau(K_L)/\tau(K^+)\simeq 2.7$
and $\theta_c$ is the Cabibbo angle.
By imposing that the value of the above expression does not exceed 
the observed value $7 \times 10^{-9}$ \cite{pdg},
we can obtain combined constraints on $|{\rm Re}(\delta_{LL})_{ds}|$, 
$\mt$ and $\sqrt{F}$, for a given $\Lambda$.
For the two extreme choices of $\Lambda$, we obtain
\bea
\label{kmm1}
& & |{\rm Re}(\delta_{LL})_{ds}| \simlt 3 \times 10^{-3}
\left( { \sqrt{F} \over \mt} \right)^8 
\left( {\mt \over 300 \, {\rm GeV} } \right)^2
\, ,  \;\;\;\;\;\; (\Lambda = \Lambda_{min})
\\
\label{kmm2}
& & |{\rm Re}(\delta_{LL})_{ds}| \simlt 6 \times 10^{-5}
\left({ \sqrt{F} \over \mt} \right)^4
\left( {\mt \over 300 \, {\rm GeV} } \right)^2
\, ,  \;\;\;\;\;\; (\Lambda = \Lambda_{max})
\eea
We now consider the terms in (\ref{fourfer}) that contribute to 
$K$-$\ov{K}$ transitions: 
\be
\label{dsds}
\cl_{eff} = -{1 \over 64 \pi^2} {\lsq \over F^4}
\left[(\mll)_{d s} \ov{d}\smub s 
- (\mrr)_{d s} d^c \smu \ov{s}^c \right]
\left[(\mll)_{d s} \ov{d}\smdb s 
- (\mrr)_{d s} d^c \smd \ov{s}^c \right] + {\rm h.c.}
\ee
Retaining again only the $LL$ contributions, we obtain:
\be
\left| {\Delta m_K \over m_K} \right|
\simeq
{ 1 \over 96 \pi^2}  
{ f_K^2 \mt^4 \lsq \over F^4}
\left| {\rm Re}(\delta_{LL})^2_{ds} \right|
\ee
where $f_K \simeq 160 \, {\rm MeV}$ is the kaon decay constant. 
We can again find constraints on the parameters by imposing that 
the value of the above expression does not exceed the experimental 
value $7 \times 10^{-15}$. 
For the two extreme choices of $\Lambda$, we obtain
\bea
\label{dk1}
& & \sqrt{ |{\rm Re}(\delta_{LL})^2_{ds}| } \simlt 5 \times 10^{-3}
\left( { \sqrt{F} \over \mt} \right)^4 
\left( {\mt \over 300 \, {\rm GeV} }  \right)
\, , \;\;\;\;\;\; (\Lambda = \Lambda_{min})
\\
\label{dk2}
& & \sqrt{ |{\rm Re}(\delta_{LL})^2_{ds}| } \simlt 7 \times 10^{-4}
\left( { \sqrt{F} \over \mt} \right)^2 
\left( {\mt \over 300 \, {\rm GeV} }  \right)
\, , \;\;\;\;\;\; (\Lambda = \Lambda_{max})
\eea

When $\sqrt{F} \gg \mt$, the above eqs.~(\ref{kmm1}-\ref{kmm2})
and (\ref{dk1}-\ref{dk2}) do not give
significant bounds on $(\delta_{LL})_{ds}$, which is
instead constrained by the non-goldstino contributions.
We recall that, in this case, the strongest bound comes 
from $\Delta m_K$ rather than $K_L \to \mu^+ \mu^-$, 
since only the former quantity receives significant 
contributions from diagrams with gluino exchange. 
When $\sqrt{F}$ 
approaches $\mt$, on the other hand, the goldstino 
contributions become more and more relevant and the 
limit on $(\delta_{LL})_{ds}$ obtained from 
$K_L \to \mu^+ \mu^-$ can be comparable or even
more stringent than that from $\Delta m_K$. 

The quantity $(\delta_{RR})_{ds}$ is constrained
in the same way as $(\delta_{LL})_{ds}$. 
The bounds from $\Delta m_K$ on the combination 
$\sqrt{|{\rm Re}[(\delta_{LL})_{ds} 
(\delta_{RR})_{ds}]|}$ are slightly more stringent
than those in eqs.~(\ref{dk1}-\ref{dk2}). 
The $\Delta S=2$ lagrangian in eq.~(\ref{dsds})
also contributes to the CP violating parameter
$\epsilon_K$. The resulting bounds on  
$\sqrt{|{\rm Im}(\delta_{LL})^2_{ds}|}$,
$\sqrt{|{\rm Im}(\delta_{RR})^2_{ds}|}$ and
$\sqrt{|{\rm Im}[(\delta_{LL})_{ds} (\delta_{RR})_{ds}]|}$
are about an order of magnitude smaller than those
of the corresponding real parts.

Flavour violating processes involving $B$ ($D$) mesons can 
be discussed along similar lines. The bounds on 
$(\delta_{LL})_{db}$ ($(\delta_{LL})_{uc}$) 
from $B_d$-$\ov{B}_d$ ($D$-$\ov{D}$) mixing,
for instance, are slightly weaker than 
the corresponding ones in the $K$-$\ov{K}$ 
system. The effective lagrangian (\ref{fourfer})
also contributes to flavour changing processes 
involving external top quarks, if the appropriate
entries in $\mll$ or $\mrr$ are non-vanishing. 
Moreover, the latter processes can also be sensitive to the 
off-diagonal entries of $\mrl$ related to the top, 
since the chiral suppression is less effective. 
In this respect, even the effective interactions due 
to tree-level sgoldstino exchange (see eq.~(\ref{fourrl}) 
and related paragraph) can play a role in decays such
as $t \to c f \bar{f}$ or $t \to c \ga \ga$.


\vspace{1 cm}

\noindent
{\large \bf 5. Flavour changing processes with external goldstinos} 
\vspace{0.3 cm}

Up to now we have discussed flavour changing processes
that have photons and ordinary fermions (leptons and quarks)
as external states, with goldstinos, sgoldstinos, sleptons,
squarks and photinos present in internal lines only. 
Now we will consider the possibility that the external
states also include goldstinos, but (again) not the other 
superpartners, which we integrate out. In particular, 
we would like to discuss whether
flavour violations can occur in low-energy processes involving 
two ordinary fermions and two goldstinos, i.e. whether
transitions such as $\mu\to e \gol \gol$ or $s\to d \gol\gol$
can take place. 

We first study what happens when sfermions and sgoldstinos
are integrated out at tree-level. Using the masses and 
couplings described in Section~2, we find that three 
types of diagrams contribute (see Fig.~5).
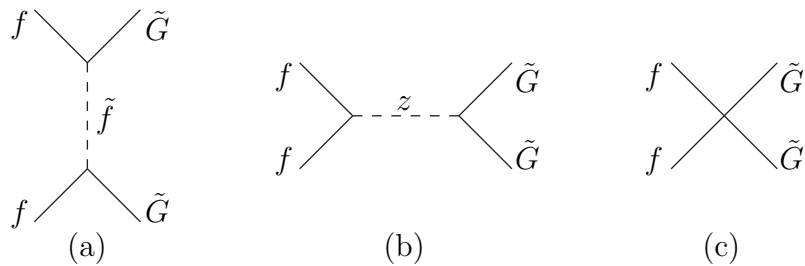
\begin{figure}[htb]
\begin{center}
\begin{picture}(320,80)(0,0)
\Line(20,0)(40,20)
\Line(40,20)(60,0)
\Line(20,80)(40,60)
\Line(40,60)(60,80)
\DashLine(40,20)(40,60){4}
\Text(20,0)[br]{$f \,$}
\Text(60,0)[bl]{$\, \gol$}
\Text(20,80)[tr]{$f \,$}
\Text(60,80)[tl]{$\, \gol$}
\Text(40,40)[l]{$ \; \ft$}
\Text(40,-10)[]{(a)}
\Line(120,20)(140,40)
\Line(120,60)(140,40)
\Line(180,40)(200,20)
\Line(180,40)(200,60)
\DashLine(140,40)(180,40){4}
\Text(120,20)[br]{$f \,$}
\Text(120,60)[tr]{$f \,$}
\Text(200,20)[bl]{$\, \gol$}
\Text(200,60)[tl]{$\,\gol$}
\Text(160,45)[]{$z$}
\Text(160,-10)[]{(b)}
\Line(280,40)(260,20)
\Line(280,40)(260,60)
\Line(280,40)(300,20)
\Line(280,40)(300,60)
\Text(260,20)[br]{$f \,$}
\Text(260,60)[tr]{$f \,$}
\Text(300,20)[bl]{$\, \gol$}
\Text(300,60)[tl]{$\, \gol$}
\Text(280,-10)[]{(c)}
\end{picture}
\end{center}
\vspace{0.3 cm}
\caption{\em Tree-level diagrams contributing to effective
interactions between two goldstinos and two matter fermions. 
The symbols $f$ and $\ft$ generically denote matter fermions 
and sfermions.} 
\label{fig5}
\end{figure}
We recall that vertices and sfermion propagators have
a non-trivial flavour structure. However, 
such structures combine in a characteristic way if we expand 
the scalar propagators around the heavy (supersymmetry breaking) 
scalar masses and treat momenta and fermion masses 
(i.e. $\Box $ and $ m m^{\dagger} $ terms) as 
perturbations. Indeed,  
we obtain\footnote{This holds for each fermion species $f$, 
of course: we omit the sum $\sum_f$ for brevity. Notice 
that generation (i.e. flavour) indices are 
still understood.}: 
\bea
\label{aaa}
(a) & \Longrightarrow &  {1 \over F^2} 
\left( \begin{array}{cc} \pzb \fb & \pz \fc \end{array} \right)
\left( \begin{array}{cc} 
\mll & \mlr \\ \mrl & \mrr   
\end{array} \right)
\left( \begin{array}{c} f\pz \\ \fcb \pzb \end{array} \right)
+ \ldots 
\\
\label{bbb}
(b) & \Longrightarrow &  - {1 \over F^2} 
\left( \begin{array}{cc} \pzb \fb & \pz \fc \end{array} \right)
\left( \begin{array}{cc} 
0 & \mlr \\ \mrl & 0   
\end{array} \right)
\left( \begin{array}{c} f\pz \\ \fcb \pzb \end{array} \right)
+ \ldots
\\
\label{ccc}
(c) & \Longrightarrow  &  - {1 \over F^2} 
\left( \begin{array}{cc} \pzb \fb & \pz \fc \end{array} \right)
\left( \begin{array}{cc} 
\mll & 0 \\ 0 & \mrr   
\end{array} \right)
\left( \begin{array}{c} f\pz \\ \fcb \pzb \end{array} \right)
\eea
where the dots denote terms suppressed by powers of momenta
or fermion masses. We immediately see that, once we sum the 
leading terms from sfermion (a) and sgoldstino (b) exchange 
with the contact term (c), a complete cancellation takes place,
as it should. The first nonvanishing contributions arise 
at the next order in the expansion, and are quadratic in momenta
or fermion masses. In particular, from sfermion 
diagrams we obtain
\bea
\label{intout}
(a) & \Longrightarrow & 
- {1 \over F^2} \left[ \pzb \fb (\Box + m^{\dagger} m)  f\pz 
+ \pz \fc (\Box + m m^{\dagger})  \fcb \pzb \right]
\nonumber
\\
& & =  - {2 \over F^2} (\pzb \fb) (\dmu f \dmd \pz) 
+ (f \to \fc)  
\eea
where we have used the equations of motion to write the second expression.
Contributions quadratic in momenta also come from sgoldstino
diagrams (b). However, these operators also contain factors like
$\mrl/m_S^2$ or $\mrl/m_P^2$, so they can be considered of 
higher order, under the assumption that $\mrl$ is chirally
suppressed\footnote{We recall that such a suppression
could be rather mild if the top quark is involved.
However, here (as before) we are mainly interested in 
low-energy processes, where only light fermions are
involved.}
(i.e. linear in fermion masses). 
So, according to this procedure, the leading non-vanishing 
interaction between on-shell goldstinos and matter fermions 
is the dimension 8 operator in eq.~(\ref{intout}). 
Notice that this operator is manifestly flavour universal
and does not depend on the superpartner spectrum. 
In particular, no trace remains 
of the flavour structure of sfermion mass matrices.

This result generalizes that obtained in the one-flavour 
case by a similar procedure \cite{asp,bfzj}. On the other hand, 
the effective low-energy interactions between goldstinos and 
matter fermions can also be obtained by other methods,
e.g. by direct non-linear realizations of supersymmetry 
\cite{volkov},
which do not require the explicit introduction of superpartners.
In such frameworks, a more general result can be obtained.
In the one-flavour case, for instance, it was recently shown 
\cite{bfzj,cllw} that the on-shell interactions between two 
goldstinos and two $f$-type fermions are described by
two independent operators:
\be
\label{nonlin}
\cl_{nl} =
-{1 \over F^2} \left[ (\pzb \smub \dnd \pz)(\fb \snub \dmd f)
- (\fb \smub \dnd \pz) C^{(f)} (\dmd \pzb \snub f) \right]
\ee
where $C^{(f)}$ is an arbitrary dimensionless coefficient\footnote{
The normalization we have chosen is such that 
$C^{(f)}= -{1 \over 4} \alpha \; \cite{bfzj} 
= {1 \over 2} C_{ff} \; \cite{cllw}$.}.
This result can be easily generalized to the multi-flavour case:
we only need to reinterpret $f$ as a collection of fermions
$f_i$ and the coefficient $C^{(f)}$ as a matrix in flavour space,
$C^{(f)}=C^{(f)}_{ij}$.
The operators for the fermions $\fc$ are analogous, with
a matrix $C^{(\fc)}$. We recall that the first operator in 
(\ref{nonlin}) corresponds to the standard coupling of 
goldstinos to the energy-momentum tensor \cite{volkov}. 
Notice that it is flavour universal, but  
differs from the operator in eq.~(\ref{intout}).
Flavour violations can only come from the second
operator in (\ref{nonlin}), if the matrix $C^{(f)}$ 
is not diagonal in the fermion mass basis.
Before discussing this possibility, it is
convenient to write eq.~(\ref{nonlin}) in two 
additional equivalent forms. Using Fierz 
rearrangements and the goldstino equations
of motion, we obtain:
\bea
\label{nl1}
\cl_{nl} & = & 
- {2 \over F^2} \left[ (\pzb \fb) (\dmu f \dmd \pz) 
+ (\dmu \pzb \fb) (\unity  - C^{(f)})(f \dmd \pz) \right] 
\\
\label{nl2}
& = & 
- {2 \over F^2} \left[ (\pzb \fb) (\dmu f \dmd \pz) 
+ {1 \over 4} (\fb (\unity  - C^{(f)}) \smub f)
\Box (\pzb \smdb \pz) \right] 
\eea
where $\unity $ is the unity matrix in flavour space. 
In particular, by comparing eq.~(\ref{nl1}) with
the interaction found by integrating out superpartners,
eq.~(\ref{intout}), we can see that the latter 
corresponds to the special case $C^{(f)} =\unity $. 
In the one-flavour case, an analogous result was 
obtained in \cite{bfzj} by explicit computation 
and comparison of scattering amplitudes.

We have seen that the non-linear formulation of
spontaneously broken supersymmetry allows for
a generic, flavour non-universal matrix $C^{(f)}$. 
On the other hand, integrating out superpartners from 
an effective theory with linearly realized supersymmetry 
has led us to find a specific, universal $C^{(f)}$. 
Therefore we can wonder whether a non-universal $C^{(f)}$ 
could emerge also in the effective linear approach, 
by generalizing the decoupling procedure and/or the 
theory itself.
We will now mention a few such possibilities. 
\begin{description}
\item[{\it (i)}] One possibility could be to keep using the structure 
described in Section~2, and then perform the decoupling of 
superpartners at one-loop level, rather than at tree-level.
Since the full computation is quite involved, we could first
focus on the quadratically divergent contributions only.  
If we do this, however, we find that the final result still has the 
form (\ref{intout}), which corresponds to a universal $C^{(f)}$. This 
follows from the fact that the inclusion of such corrections 
amounts to use a corrected K\"ahler potential. Thus, once the 
theory is expressed in terms of one-loop corrected fields, 
masses and couplings, the decoupling procedure for the
interactions under study is formally 
similar to the tree-level one.
However, this argument does not necessarily hold for 
logarithmically divergent and finite corrections, where the 
flavour structure of sfermion mass matrices might survive 
and lead to non-universal contributions to $C^{(f)}$. 
Notice that the superspace interpretation of those 
corrections corresponds to both K\"ahler and non-K\"ahler 
(i.e. higher derivative) terms.
These considerations also suggest a different (alternative)
approach, in which computations are done at the tree level, 
by starting however from an effective lagrangian which 
already contains higher derivative terms, besides the 
K\"ahlerian ones.

\item[{\it (ii)}] Another possibility could be to relax the
assumption of pure $F$-type supersymmetry breaking used
so far.
We can consider a more general situation, with larger
field content and gauge structure, such that 
non-vanishing auxiliary component vevs appear in both chiral
and vector supermultiplets (mixed $F$-$D$ breaking). 
In this case, the goldstino is a linear
combination of the fermions in those multiplets,
and the sgoldstino sector includes the bosonic 
components of such multiplets, i.e. both spin-0 
fields (like $z$ above) and spin-1 fields (see
e.g. \cite{fayet2,fayet3}). 
Sfermion masses receive both 
$F$-type and $D$-type supersymmetry breaking contributions. 
Once all this is taken into account, one can again
integrate out sfermions and sgoldstinos at tree level
and find the effective interactions between two matter 
fermions and two goldstinos. We have checked that
the leading terms again cancel, as they should 
(see also \cite{fayet3}). 
At first non-vanishing order we find dimension 8 operators 
as in eqs.~(\ref{nl1}-\ref{nl2}). Diagrams with sfermion 
exchange again give a universal contribution to
$C^{(f)}$ (the same as before).
Diagrams with spin-1 sgoldstino exchange 
give an additional model dependent contribution
to $C^{(f)}$. The latter contribution depends e.g. 
on the coupling (charges) of matter fermions
with spin-1 sgoldstinos. If these charges are
neither universal nor aligned with fermion
masses, flavour changing effects can arise.
In this case, however, one should also keep
under control the contributions of spin-1 
sgoldstino exchange to (dimension 6) flavour 
changing operators involving four matter
fermions. 

\end{description}

Exploring in more detail the possibilities mentioned
above, or other ones, lies beyond the scope of the
present paper. For the rest of our discussion,
we will rely on the fact
that supersymmetry in principle allows for the existence of 
effective (two goldstino)-(two matter fermion) operators 
with non-diagonal matrices $C^{(f)}$,
and ask what this could imply for phenomenology. 
We will see that the high dimensionality of the
effective operators implies by itself a strong
suppression, so that even low values of $\sqrt{F}$
and large flavour violating entries in  $C^{(f)}$
are allowed. 
Consider for instance the charged lepton sector, and assume
that $C_{e \mu}$ and/or $C_{e^c \mu^c}$ are non-vanishing,
so that the flavour changing decay $\mu^- \to e^- \gol \gol$
can take place. Although this decay is flavour changing, the
final state is very similar to that of the flavour conserving 
decay $\mu^- \to e^- \nu_{\mu} \ov\nu_e$, which proceeds at 
leading order through the standard Fermi interaction.
The corresponding operators can be cast into similar forms
(see eq.~(\ref{nl2})): 
\be
\label{emgg}
{1 \over 2 F^2}[C_{e \mu} (\ov{e} \smub \mu)
- C_{\mu^c e^c} (e^c \smu \ov{\mu}^c) ] 
\Box ( \pzb \smdb \pz)
\ee
\be
\label{emnn}
2 \sqrt{2} G_F (\ov{e} \smub \mu) ( \bar\nu_{\mu} \smdb \nu_e),
\ee
The presence of two derivatives in the first operator,
however, gives a strong suppression. Indeed, the ratio of 
the two decay rates scales as: 
\be
\label{ratio}
{ \Gamma(\mu^- \to e^- \gol \gol) \over
\Gamma(\mu^- \to e^- \nu_{\mu} \ov\nu_e) } \sim
(|C_{e \mu}|^2 + |C_{\mu^c e^c}|^2) 
{ m_{\mu}^4 \over G_F^2 F^4  }
\ee
Even if we take $F\simeq G^{-1}_F$ and $|C_{e \mu}|^2+|C_{e^c \mu^c}|^2
= \co (1)$, the branching ratio is tiny, $\co (10^{-13})$.
Although the features of the electron emitted with the goldstino
pair (polarization, energy and angular distributions) differ from
those of the standard channel, the numerical
suppression is so strong that detection seems impossible.
Similar considerations apply if we compare the decays 
$\tau \to e \gol \gol$ and $\tau \to \mu \gol \gol$
to the corresponding standard ones. Even though 
$m^4_{\mu}$ above is replaced by $m^4_{\tau}$, the ratios 
analogous to (\ref{ratio}) are still tiny, e.g. $\co(10^{-9})$ 
if  $F\simeq G^{-1}_F$ and the $C_{ij}$ are $\co (1)$.

We finally consider flavour changing transitions with goldstino
pair emission in the quark sector.
Consider for instance the operator 
\be
\label{kpsdgg}
{1 \over 2 F^2}[C_{s d} (\ov{s} \smub d)
- C_{d^c s^c} (s^c \smu \ov{d}^c) ] 
\Box ( \pzb \smdb \pz)
\ee
This operator does not contribute to $K^0 \to \gol \gol$,
but does contribute e.g. to $K^+ \to \pi^+ \gol \gol$. 
This is similar to the decay  $K^+ \to \pi^+ \nu \ov\nu$,
which is itself very suppressed when compared to its 
charged counterpart. For instance, in the SM one expects
(see e.g. \cite{buras})
\be
\label{kpnn}
{ \Gamma(K^+ \to \pi^+ \nu \ov\nu) \over
\Gamma( K^+ \to \pi^0 e^+ \nu_e) } \sim 10^{-9}
\ee
For $K^+ \to \pi^+ \gol \gol$, the corresponding ratio
scales as 
\be
\label{kpgg}
{ \Gamma(K^+ \to \pi^+ \gol \gol) \over
\Gamma( K^+ \to \pi^0 e^+ \nu_e) } \sim
( |C_{s d}|^2+|C_{d^c s^c}|^2) { m_K^4 \over G_F^2 F^4  }
\ee
Even in the extreme case in which $F\simeq G^{-1}_F$ and 
$|C_{d s}|$ (or $|C_{s^c d^c}|$) is $\co (1)$, 
the latter ratio is $\co (10^{-11})$, which is smaller 
than the ratio in (\ref{kpnn}).  
The rates for the analogous $B$ decays 
($B \to \pi \gol \gol$, $B \to K \gol \gol$) are
also smaller than the corresponding ones 
with neutrino pair emission. 
To weaken the effect of the low-energy suppression,
we could move to higher energies and consider 
the top quark. Operators analogous to those discussed 
above could induce, for example, the flavour changing 
decay $t \to c \gol \gol$. The corresponding rate 
would be strongly enhanced by the presence of 
$m_t$: if $F\simeq G^{-1}_F$ and $|C_{t c}|$ 
(or $|C_{t^c c^c}|$) is $\co (1)$,
$BR(t \to c \gol \gol)$ could reach
values as large as $\co(10^{-2})$.


\vspace{1 cm}

\noindent
{\large\bf 6. Summary}
\vspace{0.3 cm}

In this paper we have pointed out and discussed a new source 
of flavour violation in supersymmetric models, namely the
couplings of goldstinos with matter. Since those
couplings are strictly related to the mass spectrum
and are suppressed by the supersymmetry breaking scale
$\sqrt{F}$, significant effects on FCNC processes can be 
obtained when the following two ingredients are present: 
{(i)} the sfermion mass matrices have a non-trivial 
flavour structure in the fermion mass basis; {(ii)} 
the supersymmetry breaking scale $\sqrt{F}$ is not much 
larger than the electroweak scale.
Notice that condition {(i)} is the same feature  
that is responsible for the well-known flavour changing 
effects induced by gaugino-matter couplings. 
In the latter case, the effects are enhanced when 
the supersymmetry breaking masses $\mt$ are close to
the electroweak scale: point {(ii)} expresses
the analogous property for goldstino contributions.
In other words, for given flavour violating sfermion 
mass matrices, goldstino and non-goldstino contributions 
to FCNC become comparable when $\sqrt{F}$ and 
$\mt$ have a similar size.

These considerations especially apply to the usual
class of low-energy FCNC processes, which involve
photons, leptons and quarks in the external states. 
In this case, goldstinos (or sgoldstinos) can  
contribute as virtual particles, as well as 
sleptons, squarks and gauginos. 
In Section~3 and Section~4 we have examined the sensitivity 
of several such processes to the value of $\sqrt{F}$ 
and to the amount of flavour violation in the
sfermion masses (parametrized by the popular
quantities $\delta_{ij}$), also making comparisons 
with the conventional non-goldstino contributions. 
In Section~3 we have discussed the decay $\mu \to e \gamma$ 
as a prototype of flavour changing radiative decays. 
The analysis confirms that, when $\sqrt{F}$ and 
$\mt$ (or $M$) have a similar size, the contributions
from goldstino-slepton-photino exchange become
comparable to those from slepton-photino exchange
and can interfere either constructively 
or destructively (see Fig.~2). When goldstino 
contributions dominate, the bounds on $\delta_{e\mu}$
become stronger than the conventional ones, 
for given $(\mt,M)$ (see Fig.~3).
A similar picture has emerged  from the analysis of 
flavour violating processes with four external matter 
fermions, discussed in Section~4. 
In this case we have focused on a representative class 
of goldstino (sgoldstino) contributions, 
which we have used to obtain order-of-magnitude estimates,
taking also into account the uncertainty due to
the cutoff scale $\Lambda$.
In this context we have  first considered the decay 
$\mu\to eee$. Again, when $\sqrt{F} \sim \mt$ the 
contributions due to the goldstino multiplet 
become comparable or even more important than 
those from photino-slepton exchange.
In this situation, moreover, the decay $\mu\to eee$ seems 
to have similar or even stronger sensitivity  to 
$\delta_{e\mu}$ as compared to $\mu\to e \gamma$,
contrary to what happens in the conventional
scenario (which corresponds to $\sqrt{F} \gg \mt$). 
In the  quark sector, we have discussed  processes 
such as $K_L\to \mu^+\mu^-$ and  $K$-$\ov{K}$ transitions. 
Again, the goldstino contributions can become dominant 
for low values of $\sqrt{F}$. In this limit, the 
bounds on the parameters $\delta_{ds}$ from 
$K_L\to \mu^+\mu^-$ can be comparable or even
more stringent than those from $\Delta m_K$. 

Finally, in Section~5, we have considered processes
with two matter fermions and two goldstinos as external 
states. The corresponding four-fermion operators have 
effective dimension 8 rather than 6, due to the special 
low-energy properties of goldstinos.
When we have obtained such operators by
integrating out heavy superpartners at tree-level, 
that feature has emerged because of mutual cancellations
among the (otherwise leading) dimension 6 terms.
By using this procedure, however, we have found that
the resulting operators do not exhibit any flavour 
structure. 
The latter result is by itself quite remarkable.
On the other hand, it cannot be regarded as completely general.
Indeed, we have noted that the more general context of 
non-linearly realized supersymmetry in principle allows 
for operators with a non trivial flavour dependence. 
We have discussed how this result might be recovered
in the linear approach, e.g. by considering the case
of mixed $F$-$D$ supersymmetry breaking rather than
pure $F$-breaking.
As regards the phenomenological implications,
we have seen that, even in the presence of
flavour violating couplings, the high dimensionality
of the effective operators automatically 
suppresses potentially interesting transitions
with goldstino pair-emission. For instance, even 
in the case of maximal flavour violation and $\sqrt{F}\sim
(G_F)^{-1/2}$, meson decays such as $K \to \pi \gol \gol$ 
have smaller rates as compared to $K \to \pi \nu \ov{\nu}$.

In conclusion, our general analysis shows that
low-energy FCNC processes are sensitive
probes of supersymmetric scenarios that involve 
both a low supersymmetry breaking scale and 
some amount of flavour violation in the sfermion 
sector. It would be very interesting to see how 
both features could emerge in concrete
models at a more fundamental level.

\vspace{0.3 cm}


\vspace{1 cm}

\noindent
{\bf Acknowledgements.}
We thank F.~Feruglio, A.~Masiero and 
F.~Zwirner for discussions.
\newpage
%


\end{document}